# EEG-Inception: An Accurate and Robust End-to-End Neural Network for EEG-based Motor Imagery Classification

**Ce Zhang[1], Young-Keun Kim[2], and Azim Eskandarian[1]**

[1] Mechanical Engineering, Virginia Polytechnic Institute and State University, Blacksburg, VA, USA
[2] Mechanical and Control Engineering, Handong Global University, Pohang, Gyeongsang, South Korea

E-mail: zce@vt.edu, ykkim@handong.edu, and eskandarian@vt.edu

## Abstract

Classification of EEG-based motor imagery (MI) is a crucial non-invasive application in brain-computer interface (BCI) research. This paper proposes a novel convolutional neural network (CNN) architecture for accurate and robust EEG-based MI classification that outperforms the state-of-the-art methods. The proposed CNN model, namely EEG-Inception, is built on the backbone of the Inception-Time network, which showed to be highly efficient and accurate for time-series classification. Also, the proposed network is an end-to-end classification, as it takes the raw EEG signals as the input and does not require complex EEG signal-preprocessing. Furthermore, this paper proposes a novel data augmentation method for EEG signals to enhance the accuracy, at least by 3%, and reduce overfitting with limited BCI datasets.

The proposed model outperforms all the state-of-the-art methods by achieving the average accuracy of 88.4% and 88.6% on the 2008 BCI Competition IV 2a (four-classes) and 2b datasets (binary-classes), respectively. Furthermore, it takes less than 0.025 seconds to test a sample suitable for real-time processing. Moreover, the classification standard deviation for nine different subjects achieves the lowest value of 5.5 for the 2b dataset and 7.1 for the 2a dataset, which validates that the proposed method is highly robust. From the experiment results, it can be inferred that the EEG-Inception network exhibits a strong potential as a subject-independent classifier for EEG-based MI tasks.

## 1. Introduction

Brain-computer interface (BCI) is a direct pathway to communicate between the human brain and external devices [1]. Electroencephalography (EEG) has become one of the most common brain activity recording methods because it is non-invasive and low-cost. An important EEG-based BCI study area is motor imagery (MI), which triggers neural activities by imagining the movement of the body (e.g., left-hand and right-hand movement) [2-4]. Decoding the correct MI neural activities allows patients with motor neuron diseases (e.g., stroke, Parkinson's disease) to rehabilitate partial body movement skills with external devices' assistance. Besides body rehabilitation, EEG-based MI popular applications also include wheelchair control [5], robot arm operation [6], and quadcopter manipulation [7].

EEG-based MI signals are non-stationary, where the signal properties such as variance and mean change with time [8]. Furthermore, EEG signals have a low signal-to-noise ratio (SNR) due to numerous artifacts and noises [9]. Therefore, decoding the EEG-based MI signals requires advanced signal processing techniques and statistical learning algorithms. There have been many investigations on the EEG-based MI tasks classification, which can be categorized into conventional machine learning algorithms and deep neural network algorithms.

Conventional machine learning algorithms are composed of three steps: signal pre-processing, feature extraction, and feature classification. The signal pre-processing step's objective is to remove artifacts such as muscle/ocular movement and system noises. As for the feature extraction step, most of them are designed based on known knowledge and previous experiences about human brain dynamics and patterns. For feature classification, popular linear classifiers, such as support vector machine (SVM), and linear discriminant analysis (LDA), are applied for MI classification. Since the EEG signals' SNR is low, the extracted features that based on know knowledge are usually covered by noises and artifacts. Furthermore, most popular classification algorithms are linear, which is not suitable for non-stationary signal classification. Therefore, conventional machine learning algorithms' MI classification accuracy is usually lower than the deep learning models. Due to these drawbacks, deep learning-based neural network models are introduced to EEG-based MI classification.

The most common neural network architectures applied to the EEG-based MI research are convolutional neural network (CNN) and recurrent neural network (RNN). The CNN models employ one or several customized kernel matrices for MI features extraction at both the temporal and the spatial domains. Because of the non-linearity feature extraction and the large number of trainable parameters, a properly designed CNN models achieves higher classification accuracy than most conventional machine learning algorithms. However, the CNN models' sizes are large, which require higher computation load. Another popular neural network architecture is RNN. The RNNs use previous time-series signal output as the input, which allows the weights of each neuron are shared across time. Compared with CNN models, most RNNs are designed for extracting temporal domain features only. However, the RNN models generally requires less computation load to achieve a relatively high classification accuracy. In summary, neural network models can automatically extract and classify EEG-based MI features. Moreover, the classification accuracy is dramatically improved compared with most conventional machine learning algorithms.

However, the existing EEG-based MI neural network models still have several issues that require more extensive research. The first issue is further enhancing the classification accuracy. Even though the neural network-based classification accuracy is dramatically improved compared with conventional machine learning algorithms, most existing methods accuracies still ranges from high 70% to low 80%, which cannot be accepted in real-world applications. The second issue is the availability of EEG datasets. Since the EEG-based MI experiments on human subjects are complex processes, the available dataset size is too small. The last issue is developing a subject-independent model. Currently, EEG-based MI classifiers are subject-dependent due to variations of neuronal feedbacks among different subjects. It is necessary to develop subject-independent MI models to be readily applied to new subjects without acquisitions of training datasets.

Therefore, this paper proposes a new CNN network to increase the accuracy of MI-EEG signal classification and robustness to subject-dependency. The proposed network is called EEG-Inception and uses several inceptions [10] and residual [11, 12] modules as the backbone. Also, to tackle the limitation in the training data, a new data augmentation of EEG signals is proposed, increasing the average accuracy by 3%.

The contributions of the EEG-Inception model can be categorized as:

- A CNN-based classification achieving 88.6% in the average accuracy outperforming all other state-of-the-art methods for binary classes dataset and 88.4% for four classes dataset

- A novel data augmentation method for EEG signal to reduce overfitting and further improve classification accuracy with small training data size

- A High robustness subject-dependent dataset with a low standard deviation in classifying different subjects

- High potential for a subject-independent EEG-based MI classification

The rest of the paper is organized as follows. Section II illustrates recent related work in EEG-based MI studies. Section III explains the EEG-Inception neural network and the novel EEG data augmentation method. Section IV describes the experimental protocol of the open-source dataset that we used for evaluation. Section V presents the proposed EEG-Inception model performance and the data augmentation effectiveness results. In section VI, we conclude the proposed algorithm results and propose potential new future works based on the EEG-Inception neural network.

## 2. Related Works

As mentioned in the first chapter, EEG-based MI classification algorithms can be separated into two categories: (1) conventional machine learning algorithms and (2) neural network-based algorithms. In this chapter, we are going to discuss the details of each category.

### 1.1 Machine Learning-based Classification

For classical machine learning algorithms, the EEG signals decoding process includes signal pre-processing, feature extraction, and feature classification.

The objective of signal pre-processing is to remove noises and artifacts. Common pre-processing methods are independent component analysis (ICA) for ocular artifacts removal, canonical correlation analysis (CCA) for muscle artifacts removal, and bandpass filtering [13-15].

For MI feature extraction, it contains data analysis in time, frequency, and spatial domains. In the time domain, the EEG event-related synchronization and desynchronization (ERD/ERS) features are determined [16]. The frequency-domain analysis is often combined with the time domain through wavelet transforms (WT) or short-time Fourier transforms (STFT) [17-19]. According to Sadiq. M.T.'s work, empirical wavelet transform has also been applied for EEG-based MI classification [20, 21]. In the spatial domain, common spatial pattern (CSP), proposed by G. Pfurtscheller et al., is proved to be effective for EEG-based MI classification [22]. After the CSP algorithm had been developed, K.K. Ang has developed a new CSP-based filter bank (FBCSP) to improve the classification accuracy [23]. Based on the FBCSP, C. Zhang and A. Eskandarian have proposed a computationally efficient CSP algorithm to minimize the computing load [24]. Besides the CSP algorithm, spatial feature dimension reduction has also been proved to be an effective technique. M.T. Sadiq has applied different feature reduction algorithms such as principal component analysis (PCA), and ICA for motor imagery feature extraction [25, 26].

As for feature classification, since the input signals are thoroughly processed, traditional classifiers such as $k^{th}$ nearest neighbor (KNN), linear discriminant analysis (LDA), and support vector machine (SVM) have been proven successful for classification [27-30].

Conventional machine learning algorithms' advantages are (1) relatively simple algorithm implementation and (2) faster training period compared with neural network-based algorithms. However, since the conventional machine learning algorithms require manual feature extraction, only limited features can be extracted based on existing algorithms. Furthermore, it is known that the EEG-based MI classification performance heavily relies on feature extraction effectiveness. Therefore, the conventional machine learning algorithm classification results are relatively limited.

## 1.2. Neural Network-based Algorithms

Researchers have recently used neural network-based MI classification methods, which have shown better performance than conventional methods. The two popular neural network structures for EEG-based MI studies are convolutional neural network (CNN) and recurrent neural network (RNN).

### 1.2.1. CNN Architecture

The CNN structure can be categorized as one-dimensional CNN (1D-CNN) and multi-dimension CNN (2D/3D-CNN).

The idea of 1D-CNN is to convolve and extract the time series or frequency domain EEG signal features. Y. Li et al. proposed a novel multi-layers 1D-CNN neural network architecture (CP-MixedNet) for MI classification [31]. In their study, forty-four channels of raw EEG signals were employed as input, and multiple 1D-CNN layers were used for spatial, temporal feature extraction. The classification accuracy results indicate that the CP-MixedNet outperforms traditional machine learning algorithms such as FBCSP. Besides convolving the time-series signals, EEG frequency amplitude is also another important input. Y. R. Tabar et al. proposed taking STFT based time-frequency amplitude as input for 1D convolution and a stack auto-encoder architecture for classification [32]. They have compared the CNN combined with the auto-encoder neural network (CNN-SAE) with the CNN architecture and other state-of-the-art algorithms. The results show that the CNN-SAE network average accuracy is around 77.6%, which is better than the 1D-CNN architecture and support vector machine (SVM) classification. However, the proposed auto-encoder classifier has eight layers with the neuron number ranging from 900 to 2, which causes excessive computational load and long training periods. M. Miao et al. have developed a deep 1D-CNN architecture to extract spatial and frequency features for MI feature extraction [33]. According to their comparison, their proposed model is around 10% higher than the CSP-based feature extraction methods.

Since each 1D-CNN layer can only convolve and extract EEG features in one dimension (time, frequency, or spatial), some researchers apply 2D/3D-CNN to extract features from multiple dimensions at the same time. For EEG-based MI studies, most multi-dimension CNN structure ideas are borrowed from object detection algorithms such as AlexNet [34]. B.H. Lee et al. proposed an end-to-end CNN for multiclass MI classification [35]. Their proposed ERA-CNN model is inspired by the hierarchical deep CNN developed for visual recognition. The ERA-CNN implements multiple 2D-CNN with kernel matrix size ranges from 36 to 288. The proposed neural network achieves around 66% for seven-class MI classification. Inspired by visual recognition, some researchers have converted the EEG time series classification (TSC) problem to an image classification problem. T. Yang et al. have proposed a convolutional neural network model with a multiple dimensional kernel matrix [36]. They have converted the EEG signals into an image as input and compared the 2D and 3D kernel matrix feature extraction and classification performance. According to the experiment results, the 2D kernel matrix generally exhibits higher accuracy than the 3D kernel matrix.

### 1.2.2 RNN Architecture

For the RNN model, long-short term memory (LSTM), inspired by natural language processing, is one of the most popular EEG-based MI classification structures. Since raw EEG signals are in time series, the time domain's correlation can be used for MI class prediction. P. Wang et al. have employed the LSTM to achieve robust classification on EEG-based MI [37]. In the proposed LSTM model, the input signals are normalized raw EEG data. After one-dimensional aggregate approximation and channel selection, the EEG signals are fed into the LSTM model. The proposed algorithm number of parameters is only 746, and the average classification accuracy for Group I dataset is 76.47%. Even though the classification accuracy is not the best compared with other state-of-art algorithms, the small size of model parameters shows that the computational load is much lower than other neural networks. Besides directly feeding time-series EEG signal to the LSTM model, J. Jeong et al. designed a CNN and LSTM combination model for MI decoding and a robotic arm control [38]. The EEG signals feature are extracted through a pre-trained CNN, then the extracted features are predicted by LSTM networks. Based on their two experiments, the robotic arm moving towards correct directions success rates are 0.6 and 0.43, respectively.

Based on the previous works on neural network study, we conclude that (1) the EEG-based MI classification accuracy needs to be further improved for real-world application, and (2) an effective subject independent algorithm is still not well developed.

## 2. Proposed Network Architecture

This section introduces the architecture of the proposed EEG-Inception model and the data augmentation method. The overall workflow of the proposed EEG-Inception model is shown in Figure 1.

### 2.1. EEG-Inception Neural Network

The proposed EEG-Inception network's backbone is an ensemble of multiple inception modules and residual modules, as shown in Figure 1. The Inception module is inspired by image classification. It extracts features from both the depth (series extraction) and the width (parallel extraction). Deeper feature extraction can be achieved by adding more inception modules, while a wider feature extraction can be obtained by applying more convolutional kernel matrices in each inception module. The residual module, initially inspired by the ResNET, is used to diminish the "vanishing gradient problem" caused by deeper layers and activation functions.

The proposed network comprises the six inception modules (one initial inception modules and five intermediate inception modules) and two residual modules, as illustrated in Figure 1. Details about the inception and residual module functions are described below.

### 2.1.1 Inception Module

The proposed algorithm inception module can be categorized as "initial inception module" and "intermediate inception module." Both modules contain a bottleneck layer, multiple convolutional layers, a pooling layer, a batch normalization, and an activation function (Figure 2a and 2b). The difference between the "initial" and the "intermediate inception module" is the bottleneck layer. In this section, we explain the details about every block in the inception module accordingly.

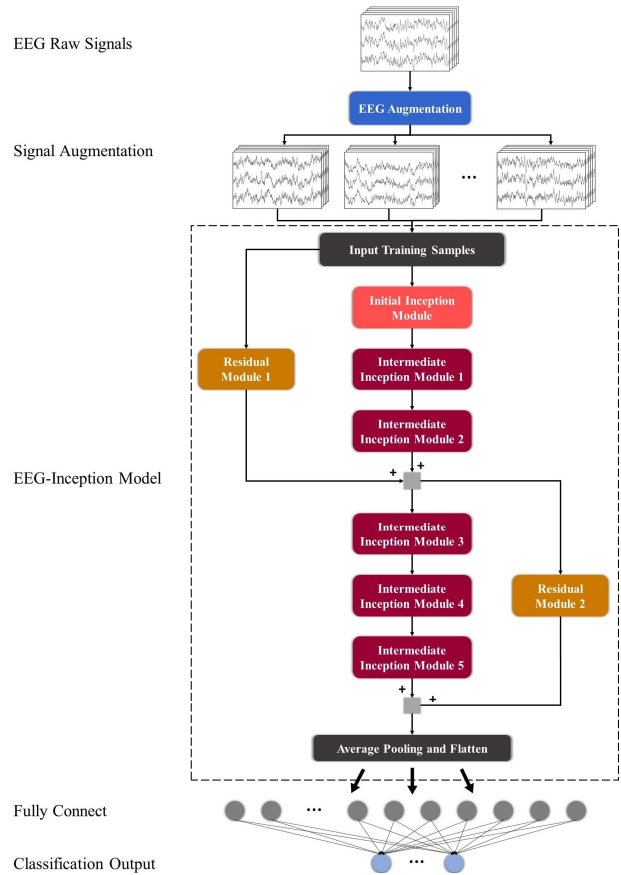

**Figure 1.** Proposed Algorithm Overall Scheme. The proposed algorithm contains a novel data augmentation method, an EEG-Inception model, and classification.

The bottleneck layer, inspired by Network in Network [39], is a [1 × 1] kernel matrix originally designed for decreasing computation load by reducing the depth of the input data dimension. It is proved to be a useful technique for feature reduction in the computer vision object detection field. However, for our initial inception module, we inverse the bottleneck layer by increasing the input data dimension from $N$ to $M$ (3 to 12 for the binary dataset, and 22 to 48 for the four-classes dataset). The reason to increase the dimension is that the provided EEG samples are one-dimensional time-series signals with a limited number of channels (three for binary classes and twenty-two for four-classes). Thus, the feature map size is too small for effective feature extraction. Therefore, the objective of inversing the bottleneck layer at the initial inception module is to increase the number of trainable weights for deeper feature extraction. For the intermediate inception module, the bottleneck layer is applied for decreasing the data dimension for lower computation load. The intermediate inception module bottleneck layer reduces the data dimension from $[4 \times M]$ to $M$. The details about the choice of the data dimension reduction size are discussed in the ablation study in the *Results and Discussions* chapter.

The convolutional layer is designed for extracting parallel features based on multiple time length. By adding the number of convolutional kernel matrices, EEG features can be extracted in parallel with various time lengths by selecting different kernel sizes. In this study, we have employed three 1D convolutional matrices for the binary dataset with kernel sizes of $[25 \times 1]$, $[75 \times 1]$, and $[125 \times 1]$, respectively (Figure 2a). For the four-classes dataset, we have used five 1D convolutional matrices with the kernel size of $[25 \times 1]$, $[75 \times 1]$, $[125 \times 1]$, $[175 \times 1]$, and $[225 \times 1]$, respectively (Figure 2b). The reason to choose the kernel sizes to be the times of twenty-five is that the provided data sample rate is 250 Hz. Thus, the extracted features from the EEG signal are based on 0.1 seconds, 0.3 seconds, and 0.5 seconds, etc. Furthermore, applying more kernel matrices for the four-classes data is because the class number is higher than the binary classes, which need more parallel features from varied time lengths to improve the classification accuracy.

The pooling layer is designed for features downsampling. In the proposed algorithm, the max-pooling is applied to select the maximum feature from the input data with a kernel size of twenty-five (0.1 seconds). After the pooling layer, a 1D convolutional layer with a $[1 \times 1]$ kernel matrix is applied to enlarge the depth of the data dimension to increase the number of learning parameters.

The batch normalization layer is a standard normalization algorithm that preventing overfitting and decreasing training steps [40].

At the end of an inception module, a ReLU activation function is employed to preserve the features of the positive values. Compared with the traditional sigmoid and the "tanh" ()" activation function, the ReLU function overcomes the "vanishing gradient problem," allowing the model to gain higher accuracy.

### 2.1.2 Residual Module

According to Figure 1, the residual module is applied after every three inception modules. The residual module is a

convolutional layer with a kernel size of $1 \times 1$, and the equation is

$$y = \mathcal{F}(x_0, \{W_i\}) + x \qquad (1)$$

where $x_0$ is the input vector, $\mathcal{F}(x_0, \{W_i\})$ represents the residual mapping from the input vector, $x$ is the output layer from the inception module, and $y$ is the summation between the residual layer output and the inception module output. K. He et. al. proves that the residual learning framework can effectively solve the learning degradation problem caused by deeper layers [34]. Thus, in our EEG-Inception module, we implement a residual module for every three inception modules to prevent learning degradation. For both residual modules kernel matrices, the input and output dimension is agreed with the inception modules' input and output for dimension consistency.

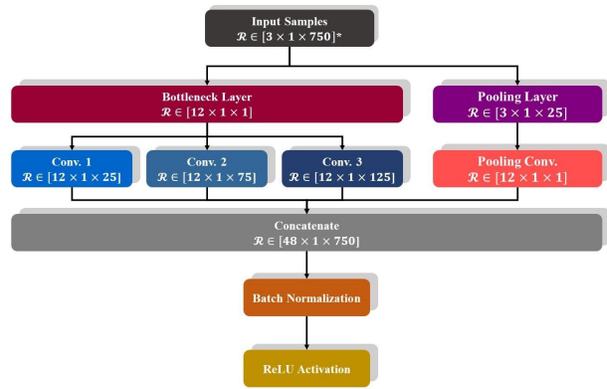

**Figure 2a.** Proposed EEG-Inception Neural Network Overall Architecture for the Binary-Classes Dataset. *The initial inception module input data depth is 3 while the intermediate inception module data depth is 48 due to concatenate from the previous inception module.

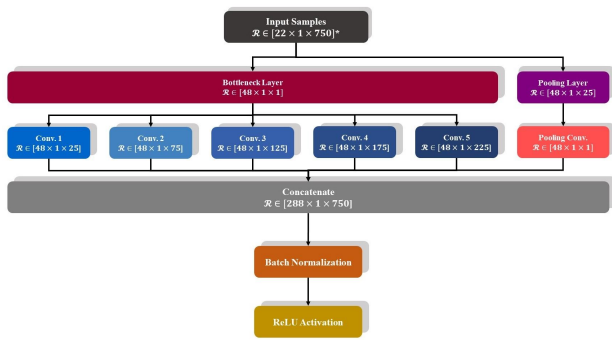

**Figure 2b.** Proposed EEG-Inception Neural Network Overall Architecture for the Four-Classes Dataset. *The initial inception module input data depth is 22 while the intermediate inception module data depth is 288 due to concatenate from the previous inception module.

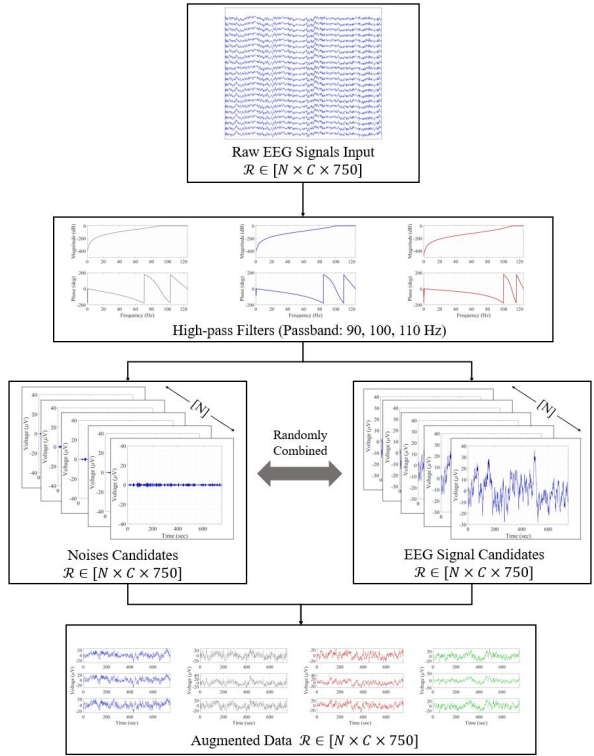

**Figure 3.** Data Augmentation Scheme. The extracted noise candidates are combined randomly with the signal candidates.

## 2.2 Data Augmentation

The objective of data augmentation is to minimize the overfitting issue by transforming the training data to extend the data size. EEG-based MI data size is usually small due to the lengthy and challenging experiment [41]. Thus, it is necessary to conduct data augmentation processing for EEG-based MI classification. Unlike standard computer vision data augmentation, EEG signals are non-stationary, which cannot be rotated, stretched, or scaled because these methods change the time-series signals properties. Therefore, noise addition is one of the best methods to expand the data size [42]. According to S. 'Muthukumaraswamy's work, most brain activities exist in a frequency range from 0-100 Hz, and the frequency above 100 Hz can be considered as artifacts and noises [43]. Thus, our proposed data augmentation method is to extract the above-100 Hz signal from one trial, then apply it to another trial, presented in Figure 3.

According to Figure 3, we have designed an $8^{th}$ order Butterworth high-pass filter with a cut-off frequency of 100 Hz to extract the noise candidates at first. Then, use the original EEG signals to subtract their noise candidates. Finally, adding noise candidates extracted from another trial, as shown in Equation 2

$$S_{aug}(i) = S_0(i) - S_n(i) + S_n(k) \qquad (2)$$

where $S_{aug}(i)$ is the augmented signal for the $i^{th}$ trial, $S_0(i)$ is the $i^{th}$ trial original signal, $S_n$ is the noise candidates, and $k$ represents a random trial number. The benefit of the proposed data augmentation method is it can increase the data size by $(n-1)$ times at most. However, increasing the data size to $n-1$ times than the original signal can cause potential signal repetitive and overfitting issues. Therefore, in our study, we increase the training data size is three times larger than the original training size for the binary dataset and six-time larger for the four-classes dataset. The augmented signal comparison with the original signal is shown in Figures 4 and 5.

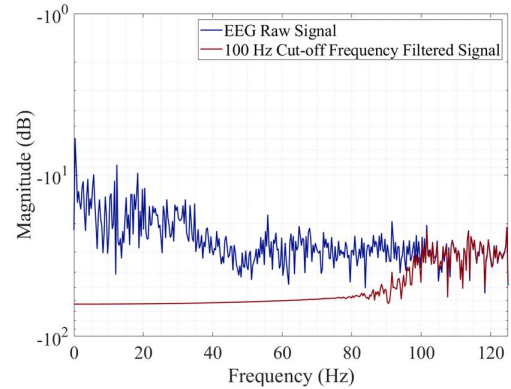

**Figure 4a.** Frequency Domain of Raw Signal and Extracted Noises. The extracted noise candidate is at a cut-off frequency of 100 Hz.

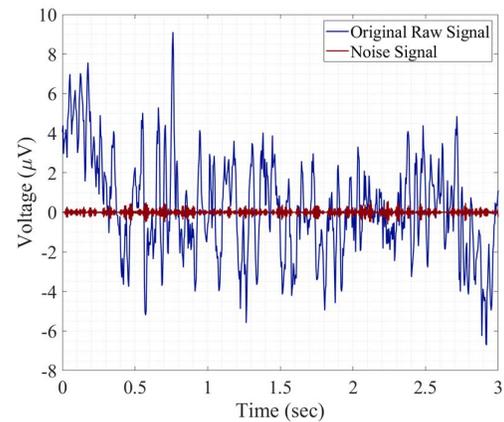

**Figure 4b.** Time Domain of Raw Signal and Extracted Noises

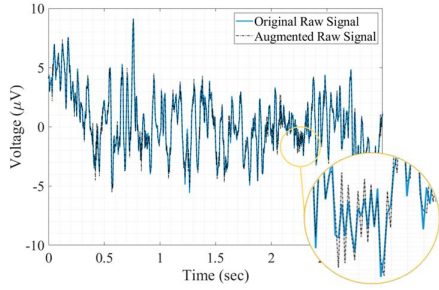

**Figure 5a.** Original and Augmented Signal for Subject 1 Trial 1

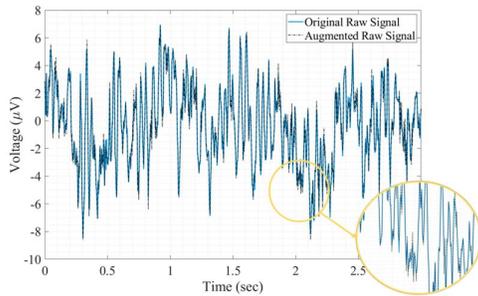

**Figure 5b.** Original and Augmented Signal for Subject 4 Trial 4

### 2.3 Ablation Study on Layer Depth

An ablation study, commonly applied in neuroscience research, is employed for artificial neural network performance analysis [44]. The objective is to investigate the classification accuracy change with different numbers of neural network layers or different features. In this study, we have tested our neural network by varying the depth of the convolutional kernel matrix. The convolutional kernel matrix dimension has varied from six to sixty-four for the binary-classes dataset and twenty-four to eighty-four for the four-classes dataset. According to our assumption, the computation time should increase because of the increasing convolutional kernel matrices depth. Besides, the classification accuracy trend is unpredictable because the larger size of trainable parameters might cause the overfitting problem.

## 3. Dataset and Experiment Protocol

### 3.1 Dataset Description

To compare the performance with other state-of-the-art algorithms, we have used two publicly available datasets from the BCI Competition IV, namely dataset 2a and dataset 2b [45]. Both datasets contain EEG and EOG signals with a sampling frequency of 250 Hz from nine subjects. For dataset 2a, subjects were required to perform four classes (left hand, right hand, feet, and tongue) MI, and for dataset 2b, subjects were asked to perform binary classes MI tasks (left hand and right hand). Figure 6 presents the localizations of the EEG channels for both datasets 2a and 2b. Twenty-two channels are provided by the dataset 2a, and three channels are provided by the dataset 2b.

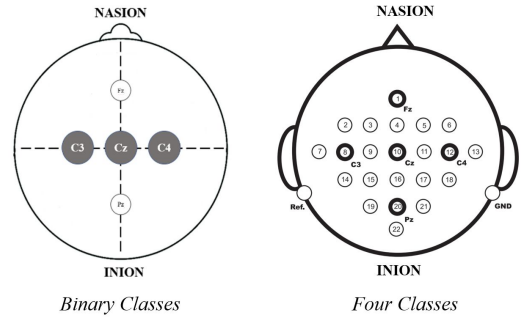

**Figure 6.** Dataset 2a and 2b Channels Localization. The 2a dataset contains 22 channels while the 2b dataset contains three channels

For the binary-classes dataset, each subject has conducted five sessions where the first two sessions are MI without results feedback (MI w/o feedback), and the last three sessions are MI with results feedback (MI w/ feedback). Figure 7 shows the paradigm of the MI w/o feedback for one trial. During the MI w/o session, the subjects are asked to perform 60 trials of MI tasks per-class, 120 trials in total. At each trial, the motor imagery period is 3 seconds. The scheme for the MI w/ feedback session of one trial is shown in Figure 8. In the MI w/ feedback session, the subjects are asked to perform 80 trials of motor imagery tasks per-class, 160 trials in total. At each trial, the motor imagery period is around 4 seconds. Moreover, the MI w/ feedback session evaluates the motor imagery performance for every trial where the green smiley face represents a correct motor imagery task while the sad red face indicates a wrong task.

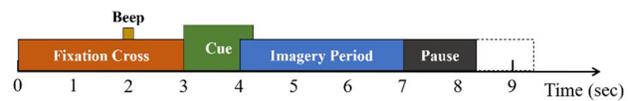

**Figure 7.** MI Binary Class w/o Feedback Session Paradigm for One Trial

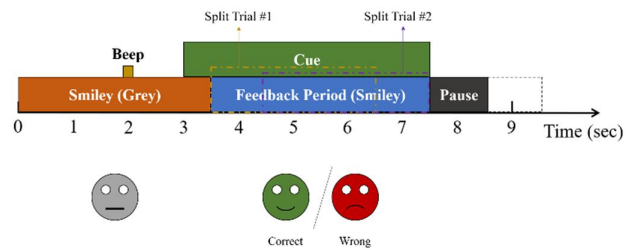

**Figure 8.** MI Binary Class w/ Feedback Session Paradigm for One Trial.

For the four classes dataset, each subject has conducted two sessions without results feedback. The experiment procedures are similar to the two classes dataset, as shown in Figure 9. The four classes dataset contains 72 trials per class (288 trials in total). At each trial, the motor imagery period is 3 seconds.

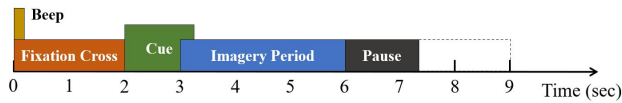

**Figure 9.** MI Four Classes Session Paradigm for One Trial

### 3.2 Experiment Protocol

Since both dataset 2a and 2b sizes are too small for neural network training and evaluation, some modifications to the input signal are necessary. For dataset 2b, we doubled the number of trials in the last three sessions by selecting the first three-second imagery period as one trial, and the last three-second imagery period is as another trial shown in Figure 8. Thus, by combining all sessions, the total number of motor imagery trials is 1200 for every subject, and the length of each trial is 3 seconds, which are used as input signals. For dataset 2a, since the motor imagery period is three seconds fixed, we only applied more augmented signals for training.

To ensure the input training data correctness, rejected trials labeled by the dataset are removed. Therefore, the total number of trials is varied based on different subjects. In the dataset, we set the training and testing data ratio around 3:1, and the details about the training and testing samples for every subject are presented in Table 1.

**Table 1.** Training/Testing Samples Summary for All Subjects at Both Binary Class and Four Classes Dataset

| | Acceptance Rate (%) | Train w/o Augmentation | Train w/ Augmentation | Test |
|---|---|---|---|---|
| *Binary Classes Dataset* | | | | |
| **S1** | 75.33% | 678 | 2034 | 226 |
| **S2** | 79.17% | 712 | 2136 | 238 |
| **S3** | 72.83% | 655 | 1965 | 219 |
| **S4** | 97.17% | 874 | 2622 | 292 |
| **S5** | 89.25% | 803 | 2409 | 268 |
| **S6** | 77.58% | 698 | 2094 | 233 |
| **S7** | 79.83% | 718 | 2154 | 240 |
| **S8** | 74.50% | 670 | 2010 | 224 |
| **S9** | 78.50% | 706 | 2118 | 236 |
| *Four Classes Dataset* | | | | |
| **S1** | 86.46% | 498 | 2988 | 140 |
| **S2** | 86.28% | 497 | 2982 | 139 |
| **S3** | 84.27% | 488 | 2928 | 137 |
| **S4** | 76.56% | 441 | 2646 | 123 |
| **S5** | 84.03% | 484 | 2904 | 135 |
| **S6** | 67.71% | 390 | 2340 | 110 |
| **S7** | 86.46% | 498 | 2988 | 138 |
| **S8** | 83.51% | 481 | 2886 | 135 |
| **S9** | 78.13% | 450 | 2700 | 127 |

## 4. Results and Discussions

The EEG-Inception model is evaluated through an open-source dataset 2a and 2b from the BCI competition IV. The proposed algorithm testing results are presented from five aspects: (1) the proposed model training process, (2) the ablation study outcome, (3) the data augmentation method effectiveness, (4) the proposed algorithm accuracy comparison with other state-of-art algorithms, and (5) a preliminary examination of the proposed EEG-Inception model for subject-independent study.

### 4.1 Training Process

The EEG-Inception neural network model is written with Python programming language-based Pytorch platform [46]. To achieve faster computation speed, the model is trained by an Nvidia GeForce RTX 2080Ti graphics card. During the training process, a backpropagation method is employed for the neural network weight update. Since the input signals are non-stationary and the model is relatively complex, we use adaptive moment estimation (Adam) as the optimizer with a general learning rate of 0.005. The Adam equation is shown in equation 3

$$\Theta_{t+1} = \theta_t - \frac{\eta}{\sqrt{\hat{v}_t} + \epsilon} \hat{m}_t \tag{3}$$

where $\theta$ is the updated parameters, $m_t$ and $v_t$ are the first and second-moment gradient, respectively, $\eta$ is the general learning rate, and $\epsilon$ is a smoothing term. The loss is calculated by the binary class cross-entropy loss function

$$H_p(q) = -\frac{1}{N}\sum_{i=1}^{N} y_i \cdot \log(p(y_i)) + (1 - y_i) \cdot \log(1 - p(y_i)) \tag{4}$$

where $y$ is the label, $p(y)$ is the predicted label probability, and $N$ is the total sample size. In our experiment, the training iteration is 100 to ensure the model is converged, and the batch size is set to be 32 for less calculation memory.

### 4.2 Ablation Study

As stated before, our ablation study is conducted by varying the convolution kernel matrix depth. Table 2 summarizes the number of model parameters and model size with different kernel matrix depths for binary-classes model and four-classes model. When increasing the layer depth, both model parameters and the model size are linearly increasing. Figure 10 presents the model classification accuracies and computation time results with varied kernel matrix depths for dataset 2b. Similar behavior can also be observed on the dataset 2a results. With increasing kernel matrix depths, the computation time follows an exponentially increasing trend while the classification accuracy is unpredictable. The classification accuracy is not guaranteed to improve by simply increasing kernel matrix depth because a deeper kernel matrix causes the model overfitting. Based on this result, the final convolution kernel matrix depth is twelve for dataset 2b and forty-eight for dataset 2a. The total model parameters are over twenty thousand for the binary classes model and over eight million for the four classes model, and the total model sizes for binary-classes and four-classes models are 10.83 and 34.10 megabytes, respectively. To better understand the proposed Inception-EEG model, the binary-classes Inception EEG model summary is tabulated in Table 3.

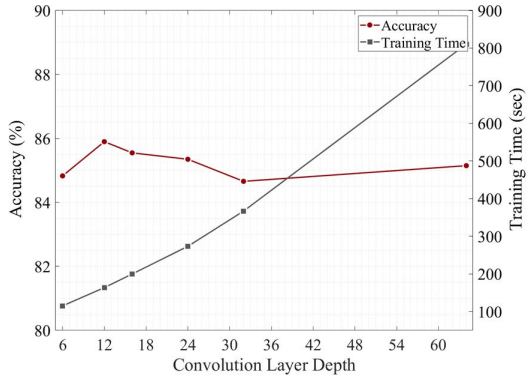

**Figure 10.** Training Time and Classification Accuracy with Different Convolution Layer Depth

**Table 2.** Model Parameters and Size Comparison with Different Layer Depth

| Binary Classes Dataset | | | Four Classes Dataset | | |
|---|---|---|---|---|---|
| Layer Depth | Total Parameters | Total Size (MB) | Layer Depth | Total Parameters | Total Size (MB) |
| 6 | 51,386 | 5.23 | 24 | 2,236,468 | 8.60 |
| 12 | 204,002 | 10.83 | 36 | 5,021,356 | 19.20 |
| 16 | 361,986 | 14.77 | 48 | 8,917,348 | 34.10 |
| 24 | 812,930 | 23.18 | 60 | 13,924,444 | 53.20 |
| 32 | 1,443,842 | 32.27 | 72 | 20,042,664 | 76.50 |
| 64 | 5,767,170 | 75.49 | 84 | 27,271,948 | 104.00 |

### 4.3 Effect of Data Augmentation

The effectiveness of the proposed data augmentation method can be evaluated from the accuracy convergence speed and the classification accuracy improvement.

For accuracy convergence speed, the proposed data augmentation method is faster than the dataset without augmentation. As shown in Figure 11, the classification accuracy with the data augmentation method is converged after 10-20 iterations, while the without augmentation method is not converged until 40-60 iterations. Since the proposed data augmentation method randomly combined the noise candidates with the signal candidates, the randomness of the training samples is improved, but the signal temporal pattern features and frequency domain features are still well preserved. Thus, with increasing the randomness of the training samples and the dataset size, the convergence speed is faster than the training without using the proposed augmentation method.

**Table 3.** EEG-Inception Neural Network Architecture Summary for Binary Classes Dataset and Four Classes Dataset

| Layer | Output Shape | Param # |
|---|---|---|
| Initial Inception Module | [48, 750] | 32,628 |
| Intermediate Inception Module_1 | [48, 750] | 33,708 |
| Intermediate Inception Module_2 | [48, 750] | 33,708 |
| Residual Module_1 | [48, 750] | 288 |
| Intermediate Inception Module_3 | [48, 750] | 33,708 |
| Intermediate Inception Module_4 | [48, 750] | 33,708 |
| Intermediate Inception Module_5 | [48, 750] | 33,708 |
| Residual Module_2 | [48, 750] | 2,448 |
| Average Pooling | [48, 750] | - |
| Linear | [1, 2 ] | 98 |
| **Initial Inception Module** | | |
| MaxPooling-1d | [3, 750] | - |
| Convolution-1d | [12, 750] | 48 |
| Convolution-1d | [12, 750] | 48 |
| Convolution-1d | [12, 750] | 3,612 |
| Convolution-1d | [12, 750] | 10,812 |
| Convolution-1d | [12, 750] | 18,012 |
| BatchNormalization-1d | [48, 750] | 96 |
| ReLU Activation | [48, 750] | - |
| **Intermediate Inception Module** | | |
| MaxPooling-1d | [3, 750] | - |
| Convolution-1d | [12, 750] | 48 |
| Convolution-1d | [12, 750] | 48 |
| Convolution-1d | [12, 750] | 3,612 |
| Convolution-1d | [12, 750] | 10,812 |
| Convolution-1d | [12, 750] | 18,012 |
| BatchNormalization-1d | [48, 750] | 96 |
| ReLU Activation | [48, 750] | - |
| **Residual Module_1** | | |
| Convolution-1d | [48, 750] | 192 |
| BatchNormalization-1d | [48, 750] | 96 |
| ReLU Activation | [48, 750] | - |
| **Residual Module_2** | | |
| Convolution-1d | [48, 750] | 2,352 |
| BatchNormalization-1d | [48, 750] | 96 |
| ReLU Activation | [48, 750] | - |
| **Parameters Summary** | | |
| Total Parameters | | 204,002 |
| Trainable Parameters | | 204,002 |

For classification results, the overall accuracy for the proposed data augmentation method is 3-4% higher than the without data augmentation training for both datasets. According to Table 4, the data augmentation method average accuracy is 2.8% higher for the binary-classes dataset and 3.6% higher for the four-classes dataset. By observing the binary-classes dataset S2-S5, and the four-classes dataset S1, S5, and S6, we find that the proposed data augmentation method can significantly improve the classification accuracy when the original accuracy is low. It is known that the main reason for low classification accuracy is artifacts caused by ocular, muscle, and other miscellaneous situations. For the "poor-performed" subjects, the noises artifacts are usually more inconsistent than the "well performed" subjects. Therefore, the proposed data augmentation method effectively extracts certain noise artifacts and randomly combined them with the MI signals, which increases the number of noisy MI samples for better model learning and higher testing classification accuracy.

**Table 4.** Classification Accuracy Comparison with Augmentation and non-Augmentation

| | *Binary Classes Dataset* | | *Four Classes Dataset* | |
|---|---|---|---|---|
| | InceptionTime w/ Aug. | InceptionTime w/o Aug. | InceptionTime w/ Aug. | InceptionTime w/o Aug. |
| S1 | 87.20% | 84.51% | 89.61% | 81.52% |
| S2 | 79.79% | 77.31% | 80.01% | 78.68% |
| S3 | 84.19% | 77.17% | 96.17% | 94.09% |
| S4 | 96.32% | 95.21% | 81.26% | 80.48% |
| S5 | 94.06% | 93.66% | 83.76% | 79.66% |
| S6 | 89.27% | 88.19% | 81.20% | 76.98% |
| S7 | 82.98% | 81.90% | 94.75% | 91.47% |
| S8 | 90.63% | 89.45% | 98.28% | 91.36% |
| S9 | 92.80% | 84.55% | 90.50% | 89.17% |
| Average | 88.58% | 85.77% | 88.39% | 84.82% |
| Std. dev. | 5.50 | 6.46 | 7.06 | 6.59 |

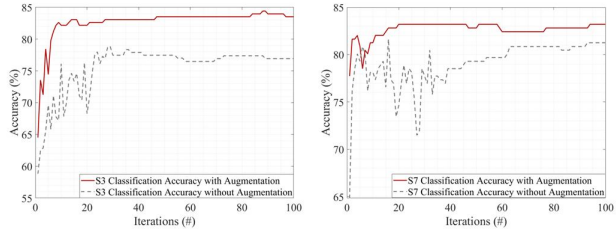

Figure 11a. Selected Subjects Binary-Classes Dataset Classification Accuracy Curve Comparison

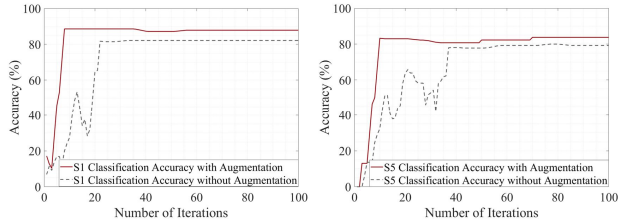

Figure 11b. Selected Subjects Four-Classes Dataset Classification Accuracy Curve Comparison

## 4.4 Comparison with State-of-the-Art Algorithms

We have compared our EEG-Inception neural network with six popular EEG-based binary motor imagery classification algorithms [47-52] for both the binary-classes dataset and four-classes dataset. To ensure a fair comparison, all the algorithms use the BCI Competition IV 2b and 2a dataset, and the results are shown in Table 5 and 6, respectively. For both datasets, we evaluate the proposed EEG-Inception network through classification accuracy, computation time, and the standard deviation of classification accuracy among all subjects. In the remaining of this section, binary-classes dataset results are discussed at first, and the four-classes dataset results are illustrated then.

### 4.4.1 Binary-Classes Dataset Results

As for average classification accuracies, the proposed EEG-Inception neural network is the highest among all state-of-art algorithms for the binary-classes dataset, which is 1% higher than the HS-CNN and almost 10% higher than the FBCSP algorithm. For single-subject classification accuracy examination, the EEG-Inception model exhibits better classification accuracies for "poorly performed" subjects. Figure 12 shows the EEG-Inception model accuracy comparison results with the top three state-of-art algorithms. According to this figure, #2 and #3 subjects exhibit the worst classification accuracies for all algorithms. The proposed EEG-Inception model maintains the classification accuracy by around 80%, which is approximately 12.5% higher than the other state-of-art algorithms. The leading causes of #2 and #3 low classification accuracies are unclear neuronal features and artifact contamination. Therefore, based on the #2 and #3 classification results, the proposed EEG-Inception algorithm can extract more effective features when subject brain signals are covered by noises and artifacts.

The computation time for the EEG-Inception network is 0.0187s for one sample. Instead of adding convolutional layers in series, the EEG-Inception network applies several convolutional layers in parallel to extract MI features from different time lengths. Therefore, the EEG-Inception network is more computationally efficient compared with other deep convolutional neural networks.

Besides the classification accuracy and computation time, we compare the average standard deviation (std. dev.) among all subjects. The proposed EEG-Inception neural network std. dev. is 5.5, which is 35.14% less than the HS-CNN algorithm and 58.31% less than the average std. dev. among all state-of-art algorithms. Such low std. dev. of the proposed algorithm indicates that the proposed algorithm is robust to all subjects, which has the potential for subject-independent learning.

### 4.4.2 Four-Classes Dataset Results

The EEG-Inception network classification accuracy for the four-classes dataset is 88.39%, which ranked 2nd among all other state-of-art algorithms [49, 51, 53-56]. The only algorithm that surpasses the EEG-Inception network for multiclass classification is the HS-CNN algorithm. The main reason that EEG-Inception is lower than the HS-CNN algorithm is because of the small training data size. Compared with the binary-classes dataset, the original four-classes dataset training sample is around 35% smaller. The EEG-Inception applies five parallel convolutional kernel matrices, which requires a more extensive training dataset to learn the kernel matrices parameters for better feature extraction performance. With the improvement of the EEG-based MI experiment protocol, creating a larger EEG-based MI dataset is feasible. Therefore, we believe that the EEG-Inception network is more suitable for larger dataset training, and the classification accuracy can be further improved.

As for computation time, the EEG-inception takes 0.0215s for testing one sample. The reason for the longer computation time than the binary-classes dataset is because the four-classes dataset contains 22 EEG channels, and the number of parallel convolutional matrices is increased from 3 to 5. Thus, the computation time is increased. However, this computation time is still fast enough for real-time processing.

The EEG-Inception network std. dev. among all subjects is 7.06 (Table 6). Similar to the classification accuracy results, the std. dev. for EEG-Inception network can be further decreased with a larger dataset. However, such low std. dev. still shows that EEG-Inception is robust to all subjects and has the potential for subject-independent learning.

**Table 5.** Binary Classes Dataset Classification Accuray Comparison with State-of-Art Methods

|  | WT-Isomap [47] | Boltzman Machine [48] | EMD-MI [49] | RSMM [50] | HS-CNN [51] | Bi-spectrum [52] | EEG-Inception |
|---|---|---|---|---|---|---|---|
| **S1** | 84.60 | 81.00 | 62.80 | 72.50 | 80.50 | 77.00 | **87.20** |
| **S2** | 66.30 | 65.00 | 67.10 | 56.40 | 70.60 | 64.50 | **79.79** |
| **S3** | 62.90 | 66.00 | **98.70** | 55.60 | 85.60 | 61.00 | 84.19 |
| **S4** | 95.80 | **98.00** | 88.40 | 97.20 | 94.60 | 96.50 | 96.32 |
| **S5** | 89.20 | 93.00 | 96.30 | 88.40 | **98.30** | 82.00 | 94.06 |
| **S6** | **97.90** | 88.00 | 75.30 | 78.70 | 86.60 | 84.50 | 89.27 |
| **S7** | 82.10 | 82.00 | 72.20 | 77.50 | 89.60 | 75.00 | 82.98 |
| **S8** | 86.30 | 94.00 | 87.80 | 91.90 | **95.60** | 91.00 | 90.63 |
| **S9** | **97.10** | 91.00 | 85.30 | 83.40 | 87.40 | 87.00 | 92.80 |
| **Average** | 84.69 | 84.22 | 81.44 | 77.96 | 87.64 | 79.83 | **88.58** |
| **Std. Dev.** | 12.72 | 11.94 | 12.74 | 14.57 | 8.48 | 11.73 | **5.50** |

### 4.5 Preliminary Subject-Independent Classification Examination

Since the EEG-Inception neural network std. dev. is the smallest compared with other state-of-art algorithms, it has the potential to be used for subject-independent classification. These results are preliminary because the EEG-Inception subject-dependent model is directly applied without further fine-tuning and adjustment. Figure 14, and Table 7 and 8 present the ROC curve and confusion matrices for the subject-independent classification for both datasets, respectively.

For the binary-classes dataset, the classification accuracy is 77.5%, which is 12.5% lower than the EEG-Inception subject dependent model but close to the RSMM and Bi-spectrum algorithms. Furthermore, according to the ROC curve, we find that the proposed data augmentation method is still effective in the subject-independent study where the AUC result is 0.03 higher than the w/o augmentation method.

For the four-classes dataset, the classification accuracy is 65.88%, which is 22% lower than the EEG-Inception subject-dependent model. Since the SoftMax equation ranges from 0 to 1, it is acceptable to see classification accuracy drop when the number of target classes increased from two to four due to the smaller probability range boundary for each class. Furthermore, the dataset size for the four-classes dataset is not large enough for subject-independent classification. Thus, the four-classes dataset subject-independent classification accuracy is lower than the binary-classes dataset subject-independent results. However, 65.88% still shows the strong potential that the EEG-Inception network can be applied for subject-independent classification with further modification.

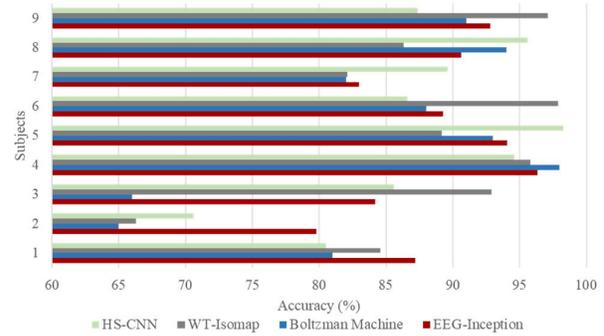

**Figure 12.** Binary Class Classification Accuracy Comparison with the Top Three State-of-Art Algorithms

**Table 6.** Four Classes Dataset Classification Accuray Comparison with State-of-Art Methods

|  | DFFS [53] | EMD-MI [49] | MEMD-Rieman [54] | Adaptive-MI [55] | R-CSP [56] | HS-CNN [51] | EEG-Inception |
|---|---|---|---|---|---|---|---|
| **s1** | 63.69 | 66.7 | **91.49** | 90.28 | 88.89 | 90.07 | 89.61 |
| **s2** | 61.97 | 63.9 | 60.56 | 54.17 | 51.39 | **80.28** | **80.01** |
| **s3** | 91.09 | 77.8 | 94.16 | 93.75 | 96.53 | **97.08** | 96.17 |
| **s4** | 61.72 | 63.2 | 76.72 | 64.58 | 70.14 | **89.66** | 81.26 |
| **s5** | 63.41 | 72.2 | 58.52 | 57.64 | 54.86 | **97.04** | 83.76 |
| **s6** | 66.11 | 70.1 | 68.52 | 65.28 | 71.53 | **87.04** | 81.2 |
| **s7** | 59.57 | 64.6 | 78.57 | 65.2 | 81.25 | 92.14 | **94.75** |
| **s8** | 62.84 | 76.4 | 97.01 | 90.97 | 93.75 | **98.51** | 98.28 |
| **s9** | 84.46 | 77.1 | 93.85 | 85.42 | **93.75** | 82.31 | 90.5 |
| **Average** | 68.32 | 70.2 | 79.93 | 73.84 | 78.01 | **91.57** | 88.39 |
| **Std. Dev.** | 11.29 | 5.92 | 14.99 | 15.72 | 17.01 | **6.49** | 7.06 |

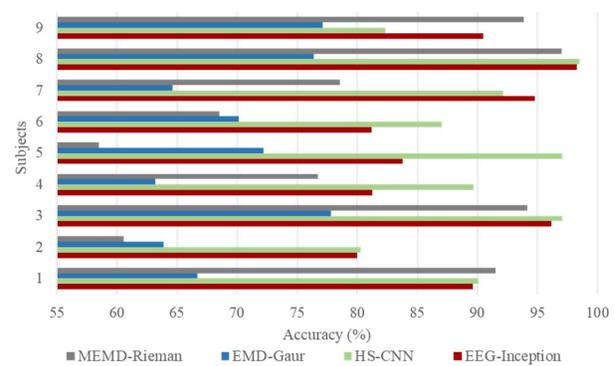

**Figure 13.** Four Classes Classification Accuracy Comparison with the Top Three State-of-Art Algorithms

independent analysis, and the classification accuracy is 77.5% and 65.9% for binary-classes and four-classes, respectively.

In the future, we will further investigate the specificity and generality of each layer from the EEG-Inception neural network. After finding the specificity and generality of the EEG-Inception network, we are going to fine-tune the EEG-Inception model so that it fits for the EEG-based MI subject independent classification task.

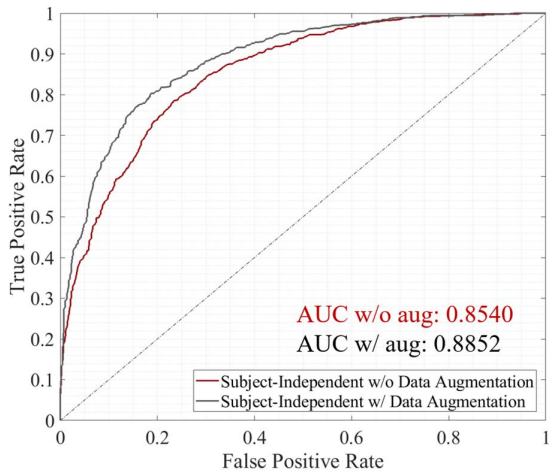

**Figure 14.** ROC Curve for Subject-Independent at Binary Classes Classification

**TABLE 7.** Confusion Matrix for Binary Classes Subject-Independent Classification

| Binary Classes Dataset Subject Independent Classification | | | |
|---|---|---|---|
| Actual \ Predicted | Left | Right | Accuracy |
| Left | 998 | 406 | 71.08% |
| Right | 85 | 687 | 88.99% |
| Avg. Accuracy | 77.44% | Kappa | 0.55 |
| F1-score | 0.737 | Recall | 0.628 |

**TABLE 8.** Confusion Matrix for Four Classes Subject-Independent Classification

| Four Classes Dataset Subject Independent Classification | | | | | |
|---|---|---|---|---|---|
| Actual \ Predict | Left | Right | Foot | Tongue | Accuracy |
| Left | 253 | 32 | 3 | 5 | 86.35% |
| Right | 60 | 214 | 6 | 12 | 73.29% |
| Foot | 79 | 57 | 159 | 16 | 51.13% |
| Tongue | 74 | 51 | 11 | 152 | 52.78% |
| Avg. Accuracy | 65.88% | | Kappa | 0.544 | |
| F1-score | 0.655 | | Recall | 0.657 | |

## 5. Conclusions and Future Works

In this paper, we presented a novel data augmentation method and modified the Inception Time neural network for EEG-based motor imagery, namely EEG-Inception. The data augmentation method can diminish the overfitting issue caused by the small data size. The proposed EEG-Inception neural network achieves an accuracy of 88.58% for the binary-classes dataset and 88.39 for the four-classes dataset. Moreover, the std. dev among all subjects is 5.5 and 7.1 for the binary-classes dataset and four-classes dataset, respectively. The low std. dev. represents the proposed algorithm features extraction is robust among all subjects. We have also conducted a preliminary examination for subject-